\documentclass[epj]{svjour}
\usepackage{graphics}
\usepackage[latin1]{inputenc}
\usepackage{bm}
\usepackage{epsfig}
\usepackage{amsfonts}
\usepackage{float}
\usepackage{amsmath,amssymb}
\usepackage{color}
\usepackage{slashed}
\usepackage{relsize}
\usepackage[toc,page]{appendix} 
\usepackage{dcolumn}
\usepackage{bm}
\newcommand{\nn}{\nonumber}
\newcommand{\be}{\begin{equation}}
\newcommand{\ee}{\end{equation}}
\newcommand{\bea}{\begin{eqnarray}}
\newcommand{\eea}{\end{eqnarray}}

\newcommand{\del}{\partial}

\begin{document}
\title{Propagation of non-linear waves in hot, ideal, and non-extensive quark-gluon plasma}
\author{Trambak Bhattacharyya\inst{1} \and Abhik Mukherjee\inst{2}
}                     
%
%
\institute{Bogoliubov Laboratory of Theoretical Physics, Joint Institute for Nuclear Research, Dubna, 141980, 
Moscow Region, Russian Federation, \email{bhattacharyya@theor.jinr.ru} \and Department of Theoretical Physics and Quantum Technology, National University of Science and Technology, MISIS, Moscow, Russian Federation,
\email{abhikmukherjeesinp15@gmail.com}}
\date{Received: date / Revised version: date}
%
\abstract{
We study the propagation of energy density perturbation in a hot, ideal quark-gluon medium in which quarks and gluons follow the Tsallis-like momentum 
distributions. We have observed that a non-extensive MIT bag equation of state obtained with the help of the quantum Tsallis-like distributions gives rise 
to a breaking wave solution of the equation dictating the evolution of energy density perturbation. However, the breaking of waves is delayed when the value of the 
Tsallis $q$ parameter and the Tsallis temperature $T$ are higher.%
} 
\titlerunning{Propagation of non-linear waves in QGP}
\authorrunning{T. Bhattacharyya and A. Mukherjee}
\maketitle

\section{Introduction}
\label{sec:intro}  

Studying the evolution of quark-gluon plasma (QGP), a hot and dense medium created in the high-energy collision experiments at the LHC, CERN and RHIC, BNL or the evolution of the high-energy particles passing through it, has been a subject matter of intense research.  To study the evolution of QGP, which is a short lived medium that expands very fast, hydrodynamic equations have been employed for a long time 
\cite{Gyulassy:1983ub,Akase:1990yd,Csernai:2003xd,Song:2008hj,Teaney:2009qa,Chaudhuri:2010in,Jaiswal:2016hex}. In addition to the vigorous change of the background medium, a bunch of high-energy 
particles (jets) passing through the bath also display sufficient modification in their distribution. These particles are formed much earlier than the formation of QGP, and if they happen to pass 
through the medium, deposit energy inside it. The evolution of their phase-space distribution has been studied microscopically using the Fokker-Planck equation \cite{mustafaplb,mooreprc,rappprc,prc} or
the Boltzmann transport equation \cite{gossiaux,grecoprc,greiner} in many articles. 

Apart from the ones mentioned above, there are other types of studies which have utilized the Boltzmann transport equation \cite{Sarwar:2015mma} or the hydrodynamic equation to investigate the evolution of the energy density perturbations in nucleus-nucleus collision \cite{rahaplb,nlwprc,nlwprd,nlwplb,tube} using, in many cases, different equations of state inspired from the MIT bag model, the mean field treatment etc. 

Energy density perturbations can originate due to high-energy particles depositing energy inside the QGP medium. The initial perturbations generated in QGP propagate through the medium showing nonlinear features and undergo modification due to medium effects. Now, if the perturbation does not break, it undergoes particle production. Otherwise, it is completely absorbed inside the bath. For example, when a high energy particle enters quark-gluon plasma, it produces energy density perturbation which may maintain its shape and is detected as the same particle. However, a lighter, low energy particle may not be able to maintain the shape of its perturbation and may be absorbed inside the medium simulating opaqueness.

Studying nonlinear waves in nuclear matter started with the derivation of Korteweg de-Vries (KdV) equation describing the propagation of baryon density pulses in proton-nucleus collisions \cite{rahaplb}.  Zero and non-zero temperature QGP is considered in \cite{nlwprc} where breaking wave solution is found using the MIT bag equation of state. This kind of nonlinear wave structures are also found in \cite{tube} for hot QGP in two spatial dimensions with cylindrical symmetry. 
However, the above calculations were done using the Boltzmann-Gibbs statistics which does not consider a fluctuating ambiance that appears in the high energy nuclear collisions.

Fluctuations of the positions of the nucleons inside the colliding nuclei may lead to fluctuation in energy density. Fluctuation inside the medium through which the perturbation travels, impacts the distribution of the particles which are created after the medium freezes-out. This can be understood from the studies of the particle spectra originating from the p-p or Pb-Pb collisions taking place in high-energy collision experiments. In the analyses of those data, experimentally obtained particle distributions have been seen to be describable by the Tsallis-like power law distributions \cite{TsCMS,TsALICE,Biro09,Cleymans13,mcdprd,TsallisTaylor,tsallisraa,Grigoryan17,bcmmp,abcp} (also called the `Tsallis non-extensive distributions') that arise in a fluctuating environment \cite{Wilk00,Wilk01,Wilk09,Biro05} which is very well the case for the QGP medium \cite{sumit}. 

In the present study, we examine the evolution of the energy density perturbation inside a hot QGP medium with the help of the Euler's equation in which the Tsallis-MIT bag equation of state \cite{TsallisMIT1,TsallisMIT2} has
been utilized to take into account the fluctuating ambiance. We observe that the non-extensive bag equation of state gives rise to a breaking wave solution for the first order energy density perturbation. However, the appearance
of the breaking waves is influenced by the factors like the Tsallis $q$ parameter, the Tsallis temperature and the parameters describing the initial condition. While carrying out our analysis, we obtain analytical closed form expressions for the Tsallis thermodynamic variables for gases comprising of very light (massless and almost massless) quantum particles (Eqs. \ref{epsilonboson}-\ref{Pfermion}). As far as our knowledge goes, these expressions were not used in any other studies, and our finding is expected to have applications in studies pertaining to thermodynamics of quantum Tsallis gas.

This article is organized as follows. In the next section, we describe the mathematical model containing some discussions about the relativistic Euler's equation, the conservation equation, the non-extensive equation of state, and the derivation of the thermodynamic variables to be used in the equation of state. Section 3 contains a detailed account of the evolution equation and its solution. Section 4 highlights the results, and we summarize as well as conclude in Section 5.


 \section{Mathematical model}
 
 Since here we derive the nonlinear evolution equation for energy density perturbation in hot QGP using Tsallis statistics and relativistic hydrodynamics, we need to evaluate the required thermodynamic variables and the equation(s) of state. As the basic dynamical equations of the system we use the relativistic Euler's equation and continuity equation for entropy density. We consider hot QGP produced at the LHC, where the baryon number density is zero at the central rapidity region and evaluate the energy density and pressure appearing in Euler's equation using the Tsallis-MIT bag equations of state.
 
\subsection{Relativistic Euler equation } 
\label{sec:RelEul}
Throughout this article we will follow the natural units {\it i.e}, $c = 1, \hbar = 1$, $k_B = 1$. As discussed in \cite{nlwprc}, the correct  description
of nonlinear wave structure in QGP including quantum effects should be given by relativistic hydrodynamics of colored superfluids which is too complicated and still in an initial stage. Admitting this fact,
 we stick to the classical relativistic hydrodynamics to study the nonlinear wave structure in hot QGP following Tsallis distribution.

The velocity field is given by, $\vec{v}=v(x,t)\hat{x}$ with a magnitude $v$ in the $x$-direction. 
We use the one-dimensional relativistic Euler's equation \cite{Weinberg,Landau} which is written as,
\bea
\frac{\partial v}{\partial t} + v \frac{\partial v}{\partial x} = \frac{v^2-1}{\epsilon+P} \left(\frac{\partial P}{\partial x}+
v \frac{\partial P}{\partial t}\right),
\label{1deuler}
\eea
where $\epsilon \equiv \epsilon(x,t)$ is the energy density and $P \equiv P(x,t)$ is the pressure. 

Our calculation will be aimed at quark-gluon plasma produced at the LHC energies where at the central rapidity region the net baryonic number density is zero, and hence we encounter vanishing chemical
potential. This leads to the fact that the particle and the anti-particle distributions are identical. 

Now, defining the entropy four-current as $s^{\mu} = s u^{\mu}$, where $u^{\mu} \equiv (\gamma, \gamma \vec{v})$ is 
the four-velocity, the continuity equation for the entropy density $s$ can be derived as (for details see \cite{nlwprc}),
\bea
v s \left(\frac{\partial v}{\partial t} + v \frac{\partial v}{\partial x}\right)
+(1-v^2)\left(\frac{\partial s}{\partial t} + s \frac{\partial v}{\partial x}+v\frac{\partial s}{\partial x}\right) = 0, \nn\\
\label{entropyconteq}
\eea
where we have used the Lorentz factor as $\gamma=1/\sqrt{1-v^2}$.

\subsection{Non-extensive MIT bag equation of state} 

\label{sec:TsEOS}

From Eqs. \eqref{1deuler} and \eqref{entropyconteq}, we observe that we have four unknown quantities appearing in two equations. Hence, we require
two additional equations which may be provided by the equations of state. In the present work we consider the non-extensive MIT bag model \cite{TsallisMIT1,TsallisMIT2} equations of state which enable us to express the pressure and the entropy density in terms of the energy density. In this model, quarks and gluons follow the quantum Tsallis fermionic (`$f$') and bosonic (`$b$') single particle distributions \cite{Hasegawa} given by,
\bea
n_f=\frac{1}{\left[1+(q-1)\frac{E_p-\mu}{T}\right]^{\frac{q}{q-1}}+1}\nn\\
n_b=\frac{1}{\left[1+(q-1)\frac{E_p-\mu}{T}\right]^{\frac{q}{q-1}}-1}
\label{TsFDBE},
\eea
where $E_p=\sqrt{p^2+m^2}$ is the single particle energy of a particle of mass $m$, $\mu$ is the chemical potential, $q$ is the Tsallis parameter, and $T$ is the Tsallis temperature. 

Tsallis single particle distributions can be obtained from the Tsallis entropy defined by \cite{Tsallis},

\bea
    S = \sum\limits_{i} \frac{p_{i}^{q}-p_{i}}{1-q},
\eea
where $p_i$ are the probabilities of the micro-states. This entropic form is non-additive, {\it i.e.} the total entropy of the system is not a summation of those of its sub-parts. This situation may appear due non-ideal plasma effects, fluctuation etc \cite{Landsberg,Tsallisbook}. Extremizing the Tsallis thermodynamic potential with respect to $p_i$ one arrives at the expression for the Tsallis probabilities and the single particle distribution \cite{pbepja}. It is shown in Ref. \cite{pbepja} that the transverse momentum spectra obtained from the single particle distribution is expressed in terms of an infinite summation and the zeroth term coincides with the phenomenological classical Tsallis distribution widely used in the literature. The forms of the Tsallis quantum distributions used in literature are similar to (but not always exactly the same as) the zeroth order approximated term when one uses the factorization approximation used in Ref. \cite{Hasegawa}. In this work we restrict ourselves within this approximated version of the Tsallis quantum distribution. We reserve the analysis with the exact distributions for future.

Assuming an ideal gas of quarks and gluons, we write down the expressions for the thermodynamic variables like 
the net baryonic number density $\rho_{\mathrm{B}}$, energy density $\epsilon$ and pressure $P$.
\begin{eqnarray}
\rho_{\mathrm{B,bag}} &=& \rho_f-\bar{\rho}_f,\label{Number}\\
\epsilon_{\mathrm{bag}} &=& \mathcal{B}+\epsilon_{b}+2\epsilon_f, \label{bagepsilon}\\
P_{\mathrm{bag}} &=& -\mathcal{B}+P_{b}+2P_f, \label{bagP}
\end{eqnarray}
where $\mathcal{B}$ is the bag constant which represents the pressure of the vacuum \cite{weisskopf}, and the subscript `$b$'(`$f$') signifies the bosonic (fermionic) contribution of the thermodynamic variables. The factor 2 in front of the fermionic parts lets us consider both particles and anti-particles. According to the bag model, quarks and gluons are assumed to stay in a spherical cavity (the `bag') in QCD vacuum and the constant $\mathcal{B}$ takes care of the confinement property.
$\rho_f$ ($\bar{\rho}_f$) signifies the particle (anti-particle) number density and as already mentioned in the previous section, the net baryonic number at the central rapidity region at the LHC energies is zero. 
Hence, we have zero baryonic chemical potential which leads to $\rho_f=\bar{\rho}_f$. 

For the energy density and pressure, we identify that the bag variables have contributions from the massless bosons (gluons) and from the massive (where mass is much less than temperature) fermions (up and down quarks) and they can be expressed in terms of $n_b$ and $n_f$ in the following way:

\begin{eqnarray}
\epsilon_{i}= g\int\frac{d^3p}{(2\pi)^3}~E_p~n_i,\label{epsilon}
P_i=g\int\frac{d^3p}{(2\pi)^3}\frac{p^{2}}{3E_p}~n_i, \label{P}
\end{eqnarray}
where $g$ is the degeneracy factor and $i=f,b$.

\subsubsection{Energy density and pressure for massless bosons}
The closed analytic expressions of the energy density and pressure of a non-extensive massless bosonic gas are given by,
\bea
\epsilon_b &=& \frac{g T^4}{2 \pi ^2 (q-1)^3 q} \left[ 3 \psi
   ^{(0)}\left(\frac{3}{q}-2\right) + \psi
   ^{(0)}\left(\frac{1}{q}\right) \right. \nn\\
&& \left.    - 3 \psi ^{(0)}\left(\frac{2}{q}-1\right) - \psi
   ^{(0)}\left(\frac{4}{q}-3\right) \right],
   \label{epsilonboson}\\
 P_b &=& \frac{g T^4}{6 \pi ^2 (q-1)^3 q} \left[3 \psi
   ^{(0)}\left(\frac{3}{q}-2\right) + \psi
   ^{(0)}\left(\frac{1}{q}\right) \right. \nn\\
   && \left. - 3 \psi ^{(0)}\left(\frac{2}{q}-1\right) - \psi
   ^{(0)}\left(\frac{4}{q}-3\right)\right], 
   \label{Pboson}
\eea
and they are related by $\epsilon_b=3P_b$. In Eqs.~\eqref{epsilonboson} and \eqref{Pboson} $\psi^{(0)}$ is the digamma function \cite{Bateman}. We defer the detailed 
calculations till the appendix of the paper.

\subsubsection{Energy density and pressure for massive fermions}
We evaluate the fermionic thermodynamic variables up to $\mathcal{O}(m^2T^2)$. We consider the plasma to be consisted of the up and down quarks having masses of 5 to 10 MeV and we have checked that for a wide range of the $q$ and $T$ parameter values appearing in the phenomenological studies of high-energy collisions, the $\mathcal{O}(m^2T^2)$ approximation works very well when the mass of the particle is around 10 MeV. In some papers it has been suggested that the $\mathcal{O}(m^2T^2)$ contribution helps one account for the non-perturbative effects \cite{pisarski} in QGP. The closed analytic expressions for the thermodynamic variables for a non-extensive gas of massive fermions up to $\mathcal{O}(m^2T^2)$ are given by the following equations:

\bea
\epsilon_f &=& \frac{g T^4}{2 \pi ^2 (q-1)^3 q} \left[3 \Phi \left(-1,1,\frac{2}{q}-1\right) \right. \nn\\
&&\left.- 3 \Phi \left(-1,1,\frac{3}{q}-2\right) \right.  \nn\\
&& \left. +\Phi
   \left(-1,1,\frac{4}{q}-3\right)-\Phi
   \left(-1,1,\frac{1}{q}\right)\right] \nn\\
   &&-\frac{g
   m^2 T^2}{4 \pi ^2 (q-1) q} \left[\Phi \left(-1,1,\frac{2}{q}-1\right)-\Phi
   \left(-1,1,\frac{1}{q}\right)\right], \nn\\
   \label{epsilonfermion}
   \\
P_f &=&\frac{g T^4}{6 \pi ^2 (q-1)^3 q} \left[3 \Phi \left(-1,1,\frac{2}{q}-1\right) \right. \nn\\
&& \left. -3 \Phi \left(-1,1,\frac{3}{q}-2\right) \right. \nn\\
&& \left. +\Phi
   \left(-1,1,\frac{4}{q}-3\right)-\Phi
   \left(-1,1,\frac{1}{q}\right)\right] \nn\\
   &&-\frac{g m^2 T^2}{4 \pi ^2 (q-1) q} \left[\Phi \left(-1,1,\frac{2}{q}-1\right)-\Phi
   \left(-1,1,\frac{1}{q}\right)\right], \nn\\
   \label{Pfermion}
\eea
where $\Phi$ is the Lerch's transcendent \cite{Bateman}. For this section also, we defer the detailed calculations till the appendix. 

\subsubsection{Calculation of $\epsilon_{\mathrm{bag}}$ and $P_{\mathrm{bag}}$}
Using Eqs.~\eqref{epsilonboson}-\eqref{Pfermion}, we write down the expressions for the 
energy density and pressure in the non-extensive bag model,

\bea
\epsilon_{\mathrm{bag}} (x,t) &=& \mathcal{B} + \epsilon_{b,1}T^4 + 2 (\epsilon_{f,1}T^4+\epsilon_{f,2}m^2T^2), \nn\\
				      &=& \mathcal{B} + \epsilon_{\mathrm{bag},1}T^4 + \epsilon_{\mathrm{bag},2}m^2T^2 \label{ebagshort} \\ \nn\\
P_{\mathrm{bag}} (x,t)          &=&  -\mathcal{B} + P_{b,1}T^4 + 2 (P_{f,1}T^4+P_{f,2}m^2T^2) \nn\\
				      &=&  -\mathcal{B} + P_{\mathrm{bag},1}T^4 + P_{\mathrm{bag},2}m^2T^2. \label{Pbagshort}
\eea

When written out long-hand, $\epsilon_{i,\ell}$ and $P_{i,\ell}$ ($i=f,b$; $\ell=1,2$) are given by,
\bea
\epsilon_{b,1} &=& \frac{g}{2 \pi ^2 (q-1)^3 q} \left[ 3 \psi
   ^{(0)}\left(\frac{3}{q}-2\right) + \psi
   ^{(0)}\left(\frac{1}{q}\right) \right. \nn\\
   && \left. - 3 \psi ^{(0)}\left(\frac{2}{q}-1\right) - \psi
   ^{(0)}\left(\frac{4}{q}-3\right) \right], \label{epsilonb1} \\
\epsilon_{f,1} &=& \frac{g}{2 \pi ^2 (q-1)^3 q} \left[3 \Phi \left(-1,1,\frac{2}{q}-1\right) \right. \nn\\
&& \left. -3 \Phi
   \left(-1,1,\frac{3}{q}-2\right)+\Phi
   \left(-1,1,\frac{4}{q}-3\right) \right. \nn\\
   && \left. -\Phi
   \left(-1,1,\frac{1}{q}\right)\right] \label{epsilonf1}, \\
\epsilon_{f,2} &=& -\frac{g}{4 \pi ^2 (q-1) q} \left[\Phi \left(-1,1,\frac{2}{q}-1\right) \right.\nn\\
&& \left.-\Phi
   \left(-1,1,\frac{1}{q}\right)\right], \label{epsilonf2} \\
P_{b,1} &=& \frac{\epsilon_{b,1}}{3},~P_{f,1} =  \frac{\epsilon_{f,1}}{3},~ P_{f,2} = \epsilon_{f,2} \label{pb1pf1pf2}. \nn\\
\eea
In terms of the above variables, $\epsilon_{\rm{bag},\ell}$ and $P_{\rm{bag},\ell}$ ($\ell=1,2$) are given by,
\bea
\epsilon_{\mathrm{bag},1} &=& \epsilon_{b,1} + 2\epsilon_{f,1}, ~P_{\mathrm{bag},1} = P_{b,1} + 2P_{f,1}, \nn\\
~\epsilon_{\mathrm{bag},2} &=& 2\epsilon_{f,2},~P_{\mathrm{bag},2} = 2P_{f,2}.
\eea

Once we get the pressure, the entropy density is given by the partial derivative of the pressure with respect to the temperature $T$, {\it i.e.} $s_{\rm{bag}}=\del P_{\rm{bag}}/\del T$. Hence we obtain,
\bea
s_{\mathrm{bag}} = 4 P_{\mathrm{bag},1}T^3 + 2 P_{\mathrm{bag},2}m^2T. \label{sbagshort}
\eea
It has been verified that in absence of the baryonic chemical potential, the pressure, energy density and the entropy density obey the following relationship,
\bea
\epsilon_{\mathrm{bag}} + P_{\mathrm{bag}} = T s_{\mathrm{bag}}.
\eea

\subsubsection{Equations of state}
In order to find out the equations of state, we express the pressure $P$ and the entropy density $s$ as functions of the energy density $\epsilon$. 
We solve Eq. \eqref{ebagshort} for the temperature $T$ in terms of the bag model energy density $\epsilon_{\rm{bag}}$, and put the solution in Eqs. \eqref{Pbagshort}, and \eqref{sbagshort} to express $P_{\rm{bag}}$, and $s_{\rm{bag}}$ as functions of  $\epsilon_{\rm{bag}}$. Solving Eq. \eqref{ebagshort} for a real and positive value of $T$ and denoting the solution with $T_{\rm{sol}}$ we obtain,
\bea
T_{\mathrm{sol}} = 
\left( \frac{-m^2 \epsilon_{\rm{bag},2} + \mathcal{R}(\epsilon_{\rm{bag}})}{2 \epsilon_{\rm{bag},2}} \right)^{\frac{1}{2}},
\eea
where $\mathcal{R}(\epsilon_{\rm{bag}}) = \sqrt{m^4 \epsilon_{\rm{bag},2}^2 + 4 \epsilon_{\rm{bag},1}(\epsilon_{\rm{bag}} - \mathcal{B})}$.
The expressions of $P_{\rm{bag}}(\epsilon_{\rm{bag}})$ and $s_{\rm{bag}}(\epsilon_{\rm{bag}})$ are given by,

\begin{eqnarray}
P_{\rm{bag}} &=& -\mathcal{B} + P_{\mathrm{bag,1}}\left( \frac{-m^2 \epsilon_{\rm{bag},2} + \mathcal{R}(\epsilon_{\rm{bag}})}{2 \epsilon_{\rm{bag},2}} \right)^{2} 
\nn\\
&& +  P_{\rm{bag,2}} m^2 \left( \frac{-m^2 \epsilon_{\rm{bag,2}}+ R(\epsilon_{\rm{bag}})}{2\epsilon_{\rm{bag,1}}} \right), \nonumber \\
s_{\rm{bag}} &=&  4 P_{\mathrm{bag,1}} 
\left( \frac{-m^2 \epsilon_{\rm{bag},2} + \mathcal{R}(\epsilon_{\rm{bag}})}{2 \epsilon_{\rm{bag},2}} \right)^{\frac{3}{2}}
\nn\\
&& +2 P_{\rm{bag,2}} m^2 \left( \frac{-m^2 \epsilon_{\rm{bag},2} + \mathcal{R}(\epsilon_{\rm{bag}})}{2 \epsilon_{\rm{bag},2}} \right)^{\frac{1}{2}}.
\label{PsEbag} 
\end{eqnarray}
\section{Nonlinear evolution equation of energy density perturbation}


In this section, we derive the evolution equation of energy density perturbation of the QGP system using the well known Reductive Perturbation Theory (RPT) \cite{RP} which helps one to deal with the perturbation which can not be neglected with respect to the mean value. For this procedure, we consider the two dynamical equations {\it i.e}, the relativistic Euler's equation (Eq.~\ref{1deuler}) and the entropy conservation equation (Eq.~\ref{entropyconteq}).
We expand the dependent variables  in terms of a perturbation parameter $\sigma$ following the RPT. Finally combining Eqs. \eqref{1deuler} and (\ref{entropyconteq}) and solving the system of equations order by order we arrive at the space-time evolution of a perturbation of the energy density in quark-gluon plasma. Now, we put $P_{\rm{bag}}(\epsilon_{\rm{bag}})$, and $s_{\rm{bag}}(\epsilon_{\rm{bag}})$ in Eqs.~\eqref{1deuler} and \eqref{entropyconteq} and in the resulting equations express ($\epsilon_{\rm{bag}}-\mathcal{B}$) in terms of temperature using Eq. \eqref{ebagshort}.
Hence we obtain,
\bea
&&\frac{\partial v}{\partial t} + v \frac{\partial v}{\partial x} + \mathcal{E}_1(1-v^2) \left( \frac{\partial \epsilon_{\mathrm{bag}}}{\partial x} + v \frac{\partial \epsilon_{\mathrm{bag}}}{\partial t} \right) 
= 0 \label{1deulerxt} \\
&& \mathcal{C}_1 (1-v^2) \left( v \frac{\partial \epsilon_{\mathrm{bag}}}{\partial x} +
\frac{\partial \epsilon_{\mathrm{bag}}}{\partial t} \right) + \mathcal{C}_2 \left(\frac{\partial v}{\partial x}+v\frac{\partial v}{\partial t} \right)
=0 \label{entropyconteqxt}, \nn\\
\eea
where $\mathcal{E}_1,~\mathcal{C}_1$ and $\mathcal{C}_2$ are given by,
\bea
\mathcal{C}_1 &=& m^2 \epsilon_{\mathrm{bag},2}+2 T^2 \epsilon_{\mathrm{bag},1}\\
\mathcal{C}_2 &=& \left(m^2 \epsilon_{\mathrm{bag},2}+\frac{2}{3} \epsilon_{\mathrm{bag},1}T^2\right) \nn\\
&&\times \left(4T^4\epsilon_{\mathrm{bag},1}+2m^2T^2
\epsilon_{\mathrm{bag},2}\right) \\
\mathcal{E}_1 &=& \frac{\left(m^2 \epsilon_{\mathrm{bag},2}+\frac{2}{3} \epsilon_{\mathrm{bag},1}T^2\right)}{\left(m^2\epsilon_{\mathrm{bag},2}+2T^2\epsilon_{\mathrm{bag},1}\right) 
\left(\frac{4}{3}T^4 \epsilon_{\mathrm{bag},1} + 2m^2T^2 \epsilon_{\mathrm{bag},2} \right)}. \nn\\
\eea

Now we express Eqs. \eqref{1deulerxt}, and \eqref{entropyconteqxt} in terms of the following dimensionless variables:
\bea
\hat{\epsilon} = \frac{\epsilon_{\rm{bag}}}{\epsilon_0}, ~ \hat{v} = \frac{v}{c_s},
\eea
where $\epsilon_0$ is a reference energy density, and $c_s$ is the velocity of sound. We express $\hat{\epsilon}$ and $\hat{v}$ in terms of a small parameter $\sigma$ following RPT as,
\bea
\hat{\epsilon} = 1+ \sigma \epsilon_1 + \sigma^2\epsilon_2 + \mathcal{O}(\sigma^3) \nonumber \\
\hat{v} = \sigma v_1 + \sigma^2 v_2 + \mathcal{O}(\sigma^3).\label{expansion}
\eea
We also change the independent variables from ($x$, $t$) to ($\xi$, $\tau$) which are related by,
\bea
\xi = \sigma^{\frac{1}{2}} \frac{(x-c_s t)}{R}, ~ \tau = \sigma^{\frac{3}{2}} \frac{c_s t}{R}, \label{stretching}
\eea
where $R$ is the characteristic length scale of the problem. 

The stretched coordinates used in Eq.~\eqref{stretching} is a part of the `reductive perturbation technique' (RPT) where the small parameter $\sigma$ signifies the smallness of the perturbed quantities relative to the corresponding equilibrium quantities. Eq.~\eqref{stretching} involves two time scales in order to explain fast dynamics at the linear limit and slower dynamics occurring at the nonlinear level. This means, at short time scale (at the lowest limit of $\sigma$) the perturbation obeys linear equation and travels with velocity $c_s$. Over a longer time scale, the wave form is influenced by nonlinearity giving rise to the breaking wave structure. 
The particular scaling with $\sigma$ used in Eq.~\eqref{stretching} comes from the idea of two time scales for long waves. The scaling must satisfy the required invariance and compatibility condition as discussed in \cite{RPT1,RPT2} in order to keep the perturbation scheme valid. A similar scaling is also used in \cite{nlwprc} for studying nonlinear waves in cold and hot QGP.

Using the expansion  (\ref{expansion}) in Eqs. \eqref{1deulerxt} and \eqref{entropyconteqxt} and equating the coefficients of different powers of $\sigma$ to zero, we get the system of differential equations for the perturbations. 

\subsection{Calculation at $\mathcal{O}(\sigma)$}
At the lowest order, {\it i.e} at $\mathcal{O}(\sigma)$ of Eqs. \eqref{1deulerxt}, \eqref{entropyconteqxt}, we get the following 
relations,
\begin{equation}
c_s^2 \frac{\partial v_1}{\partial \xi} = \epsilon_0 \mathcal{E}_1 \frac{\partial \epsilon_1}{\partial \xi}, \ \ 
\mathcal{C}_2 \frac{\partial v_1}{\partial \xi} = \epsilon_0 \mathcal{C}_1 \frac{\partial \epsilon_1}{\partial \xi}. 
\label{1o}
\end{equation}

Solving (\ref{1o}) we can determine the speed of sound $c_s$ as, 
\begin{equation}
 c_s^2 = \frac{\mathcal{C}_2 \mathcal{E}_1}{\mathcal{C}_1}. 
 \label{cs}
\end{equation}

\subsection{Calculation at $O(\sigma^2)$}

At the next order i.e, at $\mathcal{O}(\sigma^2)$ we get the nonlinear evolution equation for the energy density perturbation $\epsilon_1$.
Using the relations between $v_1$ and $\epsilon_1$ from Eq.~\eqref{1o} and Eq.~\eqref{cs} we finally get,
\bea
\frac{\partial \epsilon_1}{\partial \tau} + \frac{2 \epsilon_0 \epsilon_{\mathrm{bag},1} \epsilon_1}{3 m^4 \epsilon_{\mathrm{bag},2}^2+8 m^2 T^2 \epsilon_{\mathrm{bag},1} \epsilon_{\mathrm{bag},2}+4 T^4
   \epsilon_{\mathrm{bag},1}^2}  \frac{\partial \epsilon_1}{\partial \xi} \nn\\
   =0,
   \label{inB} \nn\\
\eea
which is the main result of our work which estimates the evolution of the first order energy density perturbation. 

Coming back to the $x$-$t$ space, we get,
\bea
&& \frac{\partial \hat{\epsilon}_1}{\partial t} + c_s \frac{\partial \hat{\epsilon}_1}{\partial x} \nn\\
&&+  \frac{2 c_s \epsilon_0 \epsilon_{\mathrm{bag},1} \hat{\epsilon}_1}{3 m^4 \epsilon_{\mathrm{bag},2}^2+
8 m^2 T^2 \epsilon_{\mathrm{bag},1} \epsilon_{\mathrm{bag},2}+4 T^4
   \epsilon_{\mathrm{bag},1}^2} \frac{\partial \hat{\epsilon}_1}{\partial x} = 0, \nn\\
   \label{inviscburg}
\eea
where $\hat{\epsilon}_1=\sigma \epsilon_1$ is the first order perturbation term in scaled energy density.
This differential equation is similar to the form obtained in Ref.~\cite{nlwprc}, differing only in the coefficient of the last non-linear term.
For a constant $T = T_{\rm{con}}$, the equation (\ref{inviscburg}) turns to be of the form of inviscid Burger's equation \cite{Burger} the details of which is discussed in the next section. If we put, $m=0, \epsilon_{\mathrm{bag},2}=0$, and $\epsilon_{\mathrm{bag},1}=37\pi^2/30$, as used in Ref.~\cite{nlwprc} for the Boltzmann-Gibbs statistics of massless fermions and bosons, we exactly get back the equation derived in there.

\subsection{ Breaking wave solutions}

For a constant $T = T_{\rm{con}}$, the coefficient of the nonlinear term of  Eq. (\ref{inB}) can be written as,
\begin{equation}
 B_c = \frac{2 \epsilon_0 \epsilon_{\mathrm{bag},1}}{3 m^4 \epsilon_{\mathrm{bag},2}^2+8 m^2 T_{\rm{con}}^2 \epsilon_{\mathrm{bag},1} \epsilon_{\mathrm{bag},2}+4 T_{\rm{con}}^4
   \epsilon_{\mathrm{bag},1}^2} .\label{Bc}
\end{equation}
Hence Eq. (\ref{inB}) becomes,
\bea
\frac{\partial \epsilon_1}{\partial \tau} + B_c \epsilon_1 \frac{\partial \epsilon_1}{\partial \xi} =0,\label{inBn}
\eea
which is nothing but the inviscid Burgers equation \cite{Burger}.  For a linear initial condition, S. Chandrasekhar found the 
explicit exact solution \cite{Chandra} of (\ref{inBn}) as,

 \bea
\epsilon_1 (\xi, \tau) = \frac{1}{B_c} \left(\frac{a \xi + b}{a \tau + 1}\right) \label{solution},
\eea   
where $a, b$ are arbitrary free parameters. Explicit solutions of Eq.~\eqref{inBn} for other relevant initial conditions are not known 
as far as our knowledge goes. In the $x$-$t$ frame the form of the exact solution (\ref{solution}) becomes,
\bea
\epsilon_1 (x, t) = \frac{1}{B_c} \left(\frac{ a \sigma^{\frac{1}{2}} (x-c_s t)  + b R}{a \sigma^{\frac{3}{2}} c_s t + R}
\right) 
\label{solutionxt},
\eea   
where $\sigma, R$ are the small parameter and characteristic length of the system respectively ( defined in the previous section) and $c_s$ is given by Eq. (\ref{cs}).
The solution (\ref{solutionxt}) behaves as a breaking wave the energy of which dissipates quickly with distance unlike solitons. 

For a more general initial condition, the solution of
\bea
\frac{\partial u(x,t)}{\partial t} + f(u) \frac{\partial u(x,t)}{\partial x} = 0,
\eea
is given by $u\left(\chi \equiv x-f(u)t\right)$ \cite{solitonbook}. 
Assuming an initial ($t=0$) energy perturbation distribution of
\bea
\hat{\epsilon}_1 (x) = A ~\mathrm{Sech}^2\left(\frac{x}{B}\right),
\eea
the solution of Eq.~\eqref{inviscburg} can be written as,
\bea
\hat{\epsilon}_1 (\chi) = A ~\mathrm{Sech}^2\left(\frac{\chi}{B}\right),
\eea
where 
\bea
\chi = x-c_s(1+B_c \hat{\epsilon}_1 (\chi))t
\eea

\begin{figure*}[!htb]
\vspace*{+1cm}
\minipage{0.39\textwidth}
\includegraphics[width=\linewidth]{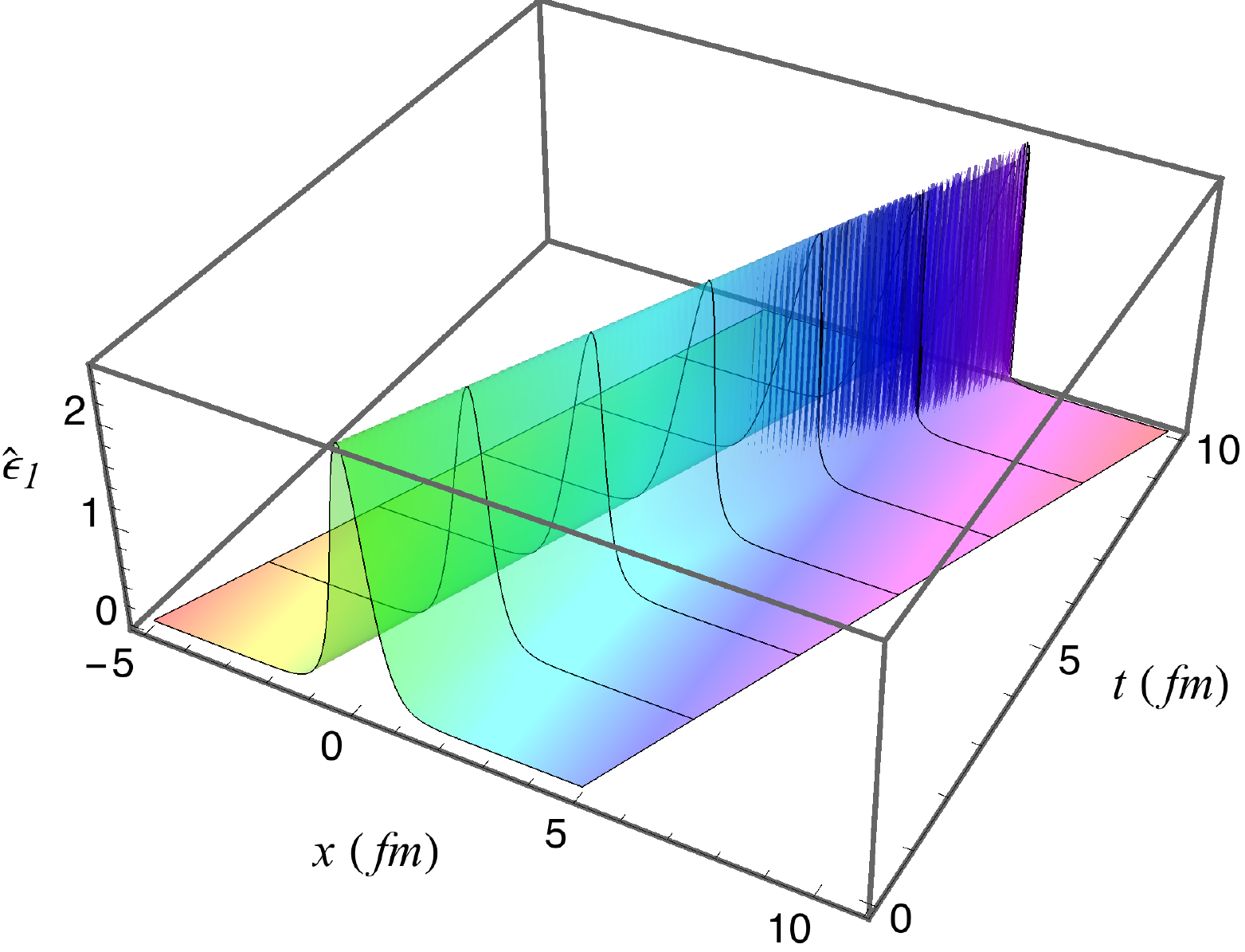}
\caption{Evolution of the energy density perturbation with time $t$ and space coordinate $x$. $A=2.5$, $B=0.5$ $fm$, $q=1.08$, and $T=140$ MeV.}
\label{figsol3d}
\endminipage\hfill
\minipage{0.46\textwidth}
\vspace*{-0cm}
\hspace*{-0cm}
\includegraphics[width=\linewidth]{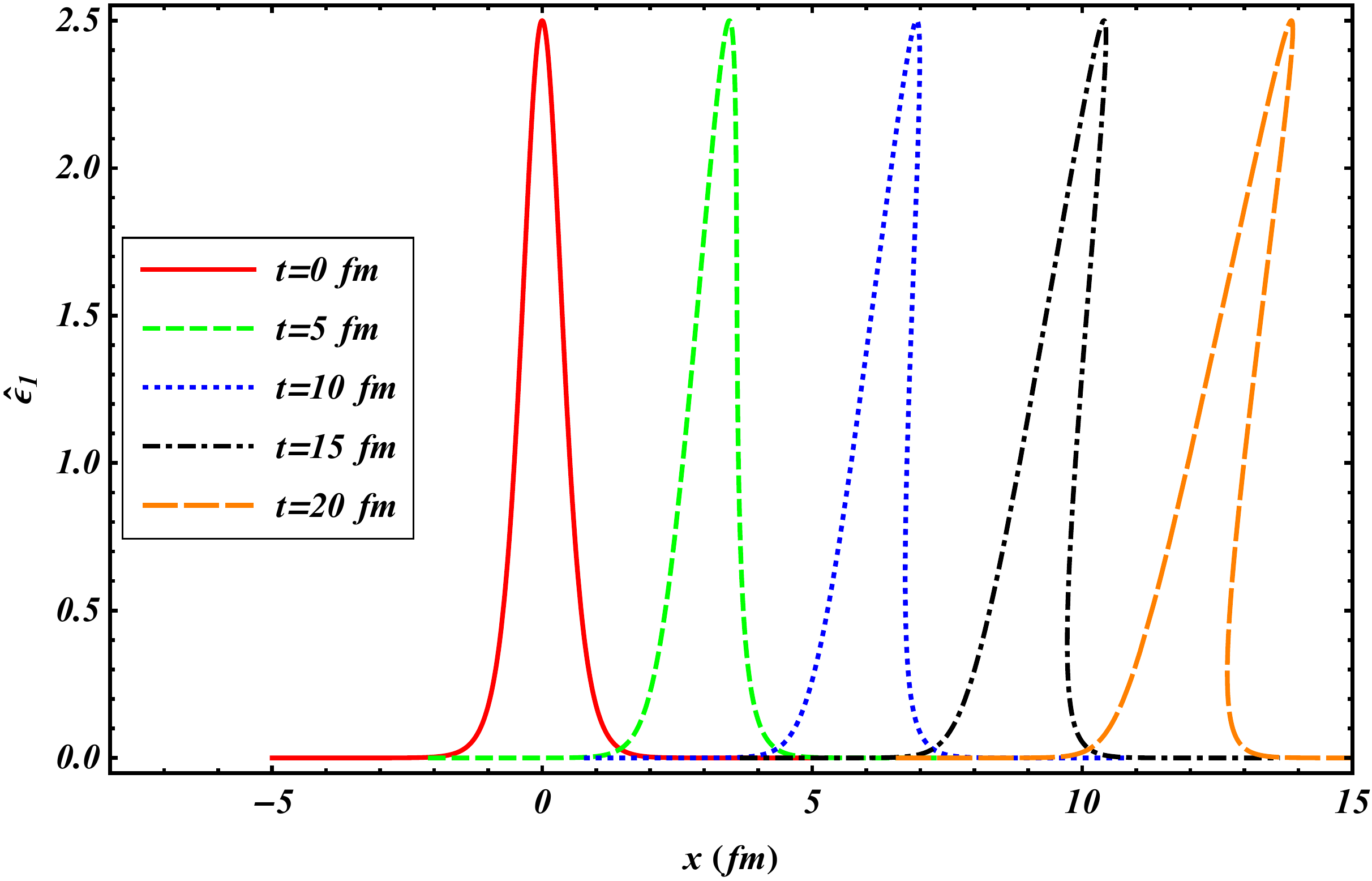}
\caption{Solutions of Fig.~\ref{figsol3d} at different times on a two-dimensional plot. $A=2.5$, $B=0.5$ $fm$, $q=1.08$, and $T=140$ MeV.}
\label{figsol2d}
\endminipage\hfill
\end{figure*}

\begin{figure*}[!htb]
\vspace*{+1cm}
\minipage{0.46\textwidth}
\vspace*{-0cm}
\hspace*{-0cm}
\includegraphics[width=\linewidth]{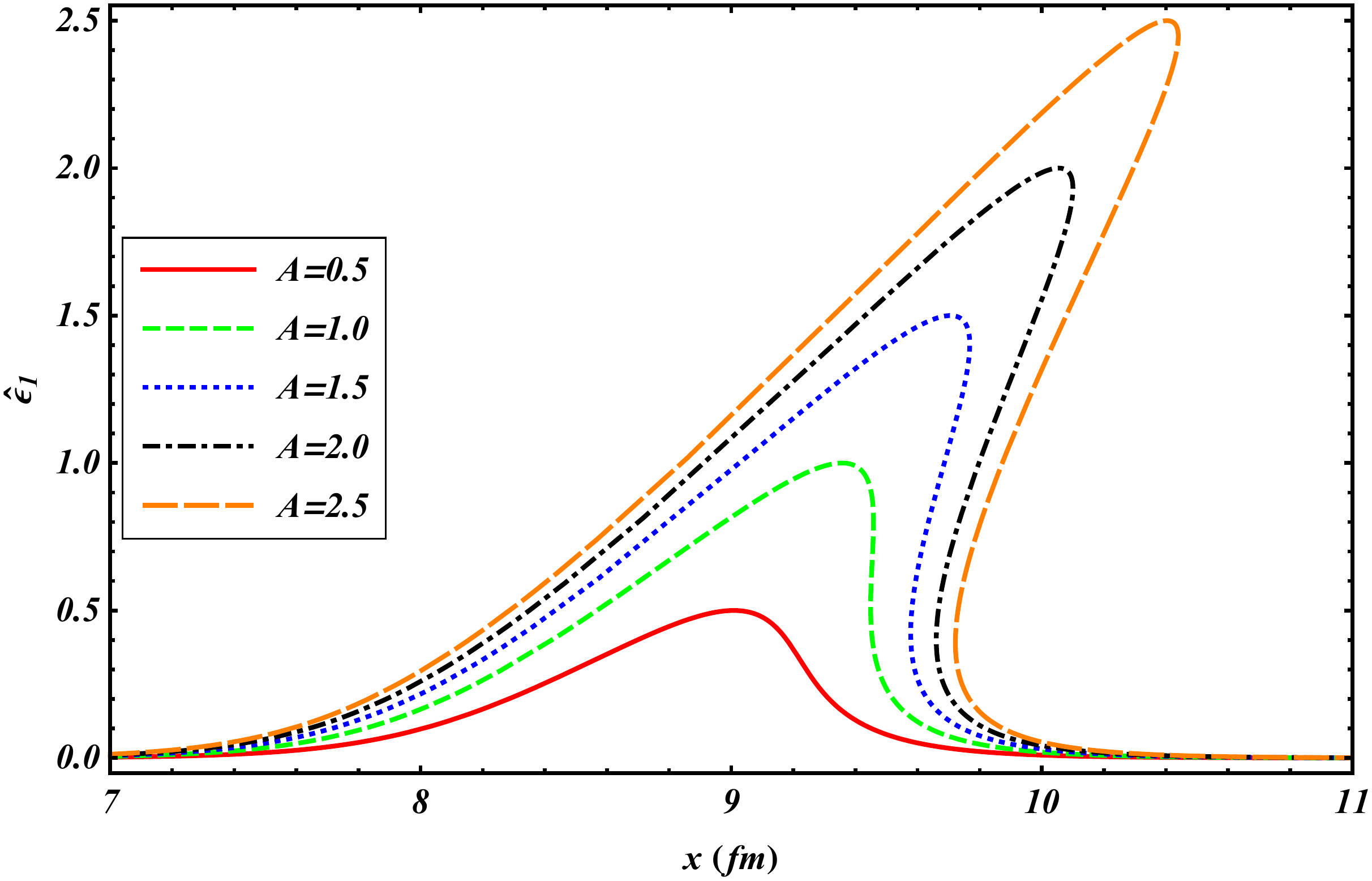}
\caption{Solutions at $t=15$ $fm$ for different amplitudes of the initial wave. $B=0.5$ $fm$, $q=1.08$, and $T=140$ MeV.}
\label{figsolvsA}
\endminipage\hfill
\minipage{0.46\textwidth}
\vspace*{-0cm}
\hspace*{-0cm}
\includegraphics[width=\linewidth]{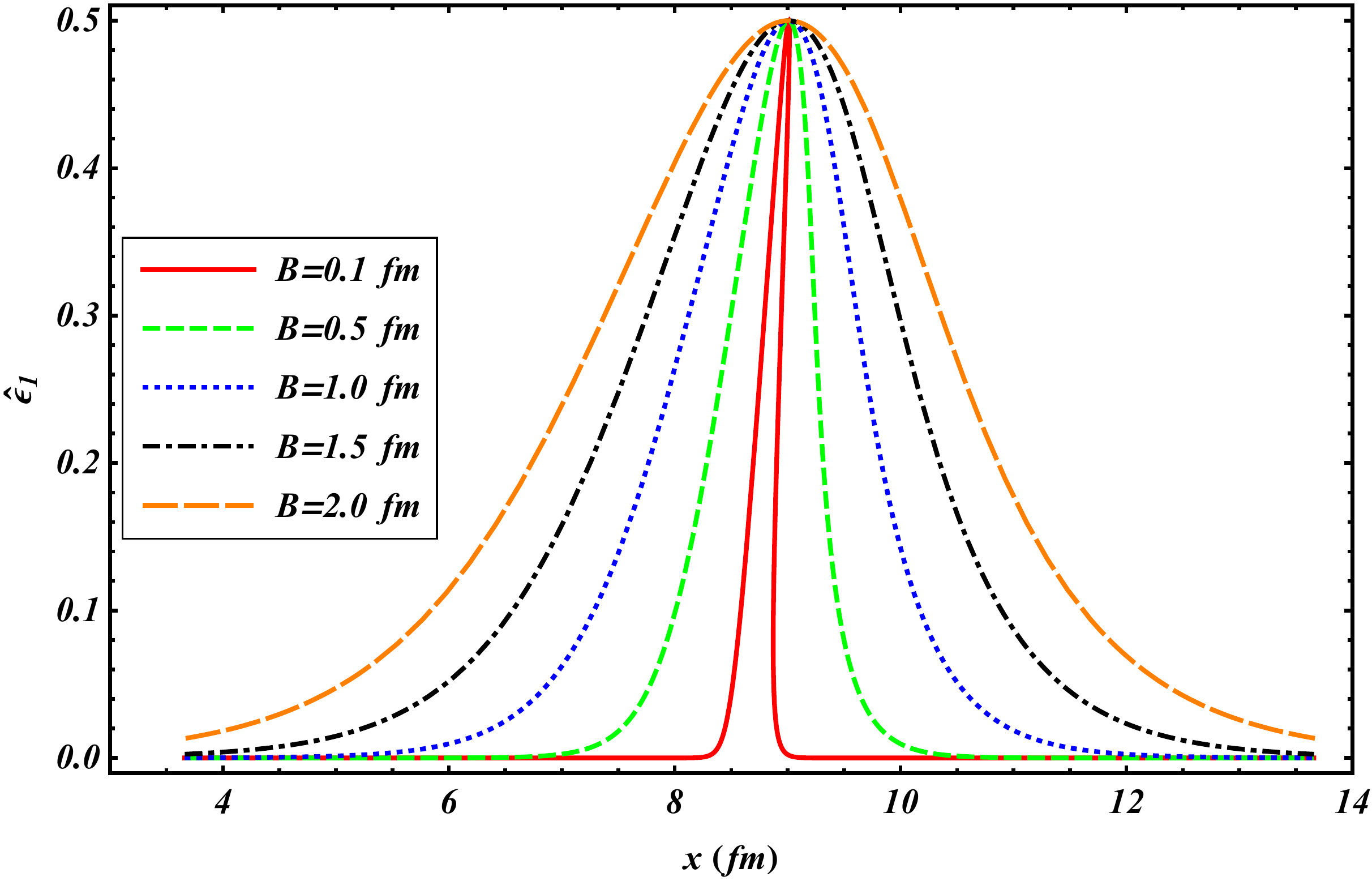}
\caption{Solutions at $t=15$ $fm$ for different widths of the initial wave. $A=0.5$, $q=1.08$, and $T=140$ MeV.}
\label{figsolvsB}
\endminipage\hfill
\end{figure*}

\begin{figure*}[!htb]
\vspace*{+1cm}
\minipage{0.46\textwidth}
\includegraphics[width=\linewidth]{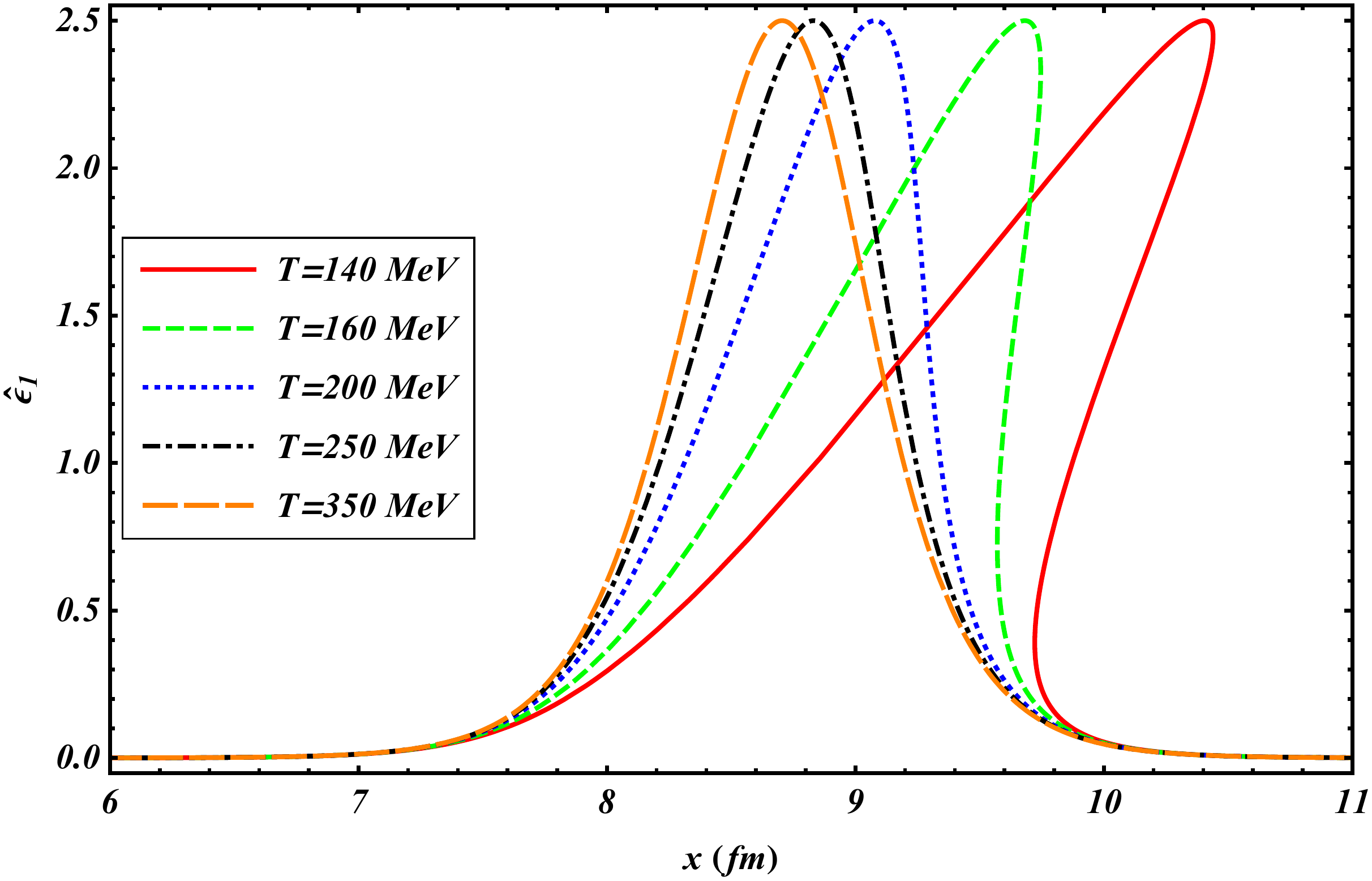}
\caption{Solutions at $t=15$ $fm$ for different Tsallis temperature values. $A=2.5$, $B=0.5$ $fm$, and $q=1.08$.}
\label{figsolvsT}
\endminipage\hfill
\minipage{0.46\textwidth}
\vspace*{-0cm}
\hspace*{-0cm}
\includegraphics[width=\linewidth]{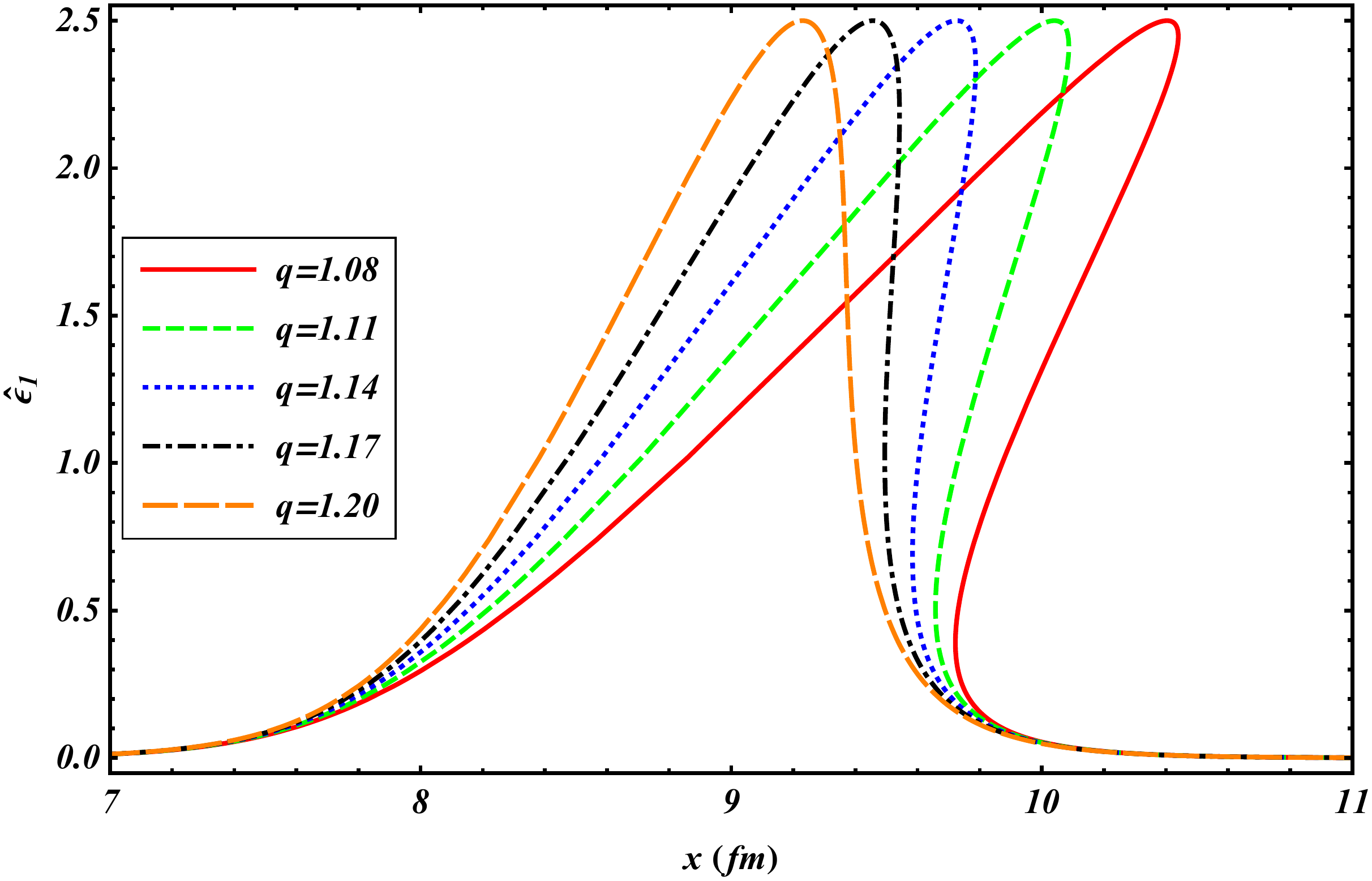}
\caption{Solutions at $t=15$ $fm$ for different Tsallis parameter values. $A=2.5$, $B=0.5$ $fm$, and $T=140$ MeV.}
\label{figsolvsq}
\endminipage\hfill
\end{figure*}
\section{Results and discussion}
For a fixed value of the reference energy density ($\epsilon_0=0.16$ GeV $fm^{-3}$) and mass $m=10$ MeV (which is of the order of the down quark mass),
the final solution depends on the amplitude $A$ of the initial wave, its width $B$, the Tsallis entropic parameter $q$, and the Tsallis temperature $T$. 

In Fig.~\eqref{figsol3d}, we plot the solutions of Eq.~\eqref{inviscburg} on the $x$-$t$ plane. For the sake of a better understanding, we present Fig.~\eqref{figsol2d} which
displays a two-dimensional version of Fig.~\eqref{figsol3d}.  For both Figs.~\eqref{figsol3d} and \eqref{figsol2d}, 
$A=2.5$, $B=0.5$ $fm$, $q=1.08$, and $T=140$ MeV. It is observed from Fig.~\eqref{figsol2d} that at around $t=10$ $fm$, the solutions start
becoming multiple-valued functions thereby implying the breaking of the waves. 

The solutions, however, depend on the width and the amplitude of the waves as well. It is apparent from Fig.~\eqref{figsolvsA}, that beyond $A=1$ (when $t=15$ $fm$, $B=0.5$ $fm$, $q=1.08$, and $T=140$ MeV), 
the solutions start becoming multiple-valued. This means, the more the amplitude of the initial energy density perturbation is, the more it becomes prone to be a breaking wave. On the other hand, a narrower 
initial energy density perturbation is more likely to give rise to a breaking wave solution (as seen in Fig.~\ref{figsolvsB}).

The temperature $T$, and the Tsallis parameter $q$ also influence the solutions so that moving from a lower to a higher temperature value makes 
functions single-valued. Hence, the lower the Tsallis temperature is, the more likely it is to get a breaking wave solution (as seen in Fig.~\ref{figsolvsT}, where 
$t=15$ $fm$, $A=2.5$, $B=0.5$ $fm$, and $q=1.08$). Similarly, a lower value of the $q$ parameter indicates a higher likelihood of obtaining breaking 
waves (as seen in Fig.~\ref{figsolvsq}, where $t=15$ $fm$, $A=2.5$, $B=0.5$ $fm$, and $T=140$ MeV).

\section{Summary and Conclusions}
In summary, we have studied the evolution of the first order energy density perturbation in hot, ideal, and non-extensive quark-gluon plasma. To take into 
account the fluctuating ambiance, we have used a non-extensive version of the MIT bag model and find a breaking wave solution. However, we observe 
that the breaking of the energy density perturbation is influenced by the temperature and the Tsallis $q$ parameter which according to Ref. \cite{Wilk00}
is related to the relative variance in a scale variable {\it i.e.} temperature. We observe that the breaking is favoured by decreasing values of both temperature
and the $q$ parameter. This may have implications in the LHC phenomenology as the quark-gluon plasma medium formed in this region may have 
higher initial temperature and $q$ value in comparison with that formed at the experiments having lower beam energies. So, the resulting wave may behave as a stable one even after a long time.

We have also verified that the at a particular instant a wave with higher amplitude and a smaller width is more prone to breaking which is intuitively understandable. During the analysis we have also provided the closed analytical formulae for the Tsallis thermodynamic variables of an ideal gas of massless bosons and very light fermions. This can be seen as an extension of an earlier work for the classical Tsallis distribution by one of us \cite{bcmprd}. 
These results can be used in the studies pertaining to the thermodynamics of quantum Tsallis gases. However, an unapproximated expression for the thermodynamic variables may be used in the future studies to examine whether that affects the present conclusions. In addition to that, the present work relies on some simplifications like one-dimensional expansion. An extension of the present work considering expansion in higher dimensions should be done.  
Within the domain of the Boltzmann-Gibbs statistics, whether one obtains a breaking wave or a localized wave depends on the equation of state. For example, in \cite{nlwprd} an equation of state inspired by the mean field QCD model resulted in the Korteveg-de Vries (KdV) soliton. But, no similar study has been reported in the field of the Tsallis statistics. It will also be interesting to explore this problem
. 
We reserve these studies for future.

\begin{acknowledgement}
T. B. acknowledges partial support from the joint projects between the JINR and IFIN-HH. The authors acknowledge Prof. Jan-e Alam and Mr. 
M. Rahman for fruitful discussions.
\end{acknowledgement}

\begin{center}
\section*{Appendix}
\end{center}
\begin{appendix}
\section{Pressure for the massless case}
The pressure of a gas of massless fermions is given by,
\bea
P= \frac{g}{6\pi^2} \int_0^{\infty} dp \frac{p^3}{\left[\left(1+(q-1)\frac{p}{T}\right)^{\frac{q}{q-1}}+1\right]}.
\eea
The Tsallis Fermi-Dirac single particle distribution is now expressed in terms of the summation of an infinite series
involving different powers of the Tsallis Maxwell-Boltzmann distributions. Now, swapping the integral and the summation 
we obtain the following expression for the pressure,
\bea
P &=& \frac{g}{6\pi^2} \sum_{s=1}^{\infty} (-1)^{s+1}\int_0^{\infty} dp \frac{p^3}{\left(1+(q-1)\frac{p}{T}\right)^{\frac{sq}{q-1}}}. \nn\\
&=& \frac{g T^4}{6\pi^2} \sum_{s=1}^{\infty} (-1)^{s+1} \frac{\Gamma\left[\frac{sq}{q-1}-4\right]}{(q-1)^4 \Gamma\left[\frac{sq}{q-1}\right]} \nn\\
&=& \frac{g T^4}{6 \pi ^2 (q-1)^3 q} \left[3 \Phi \left(-1,1,\frac{2}{q}-1\right) \right. \nn\\
&& \left. -3 \Phi
   \left(-1,1,\frac{3}{q}-2\right)+\Phi
   \left(-1,1,\frac{4}{q}-3\right) \right. \nn\\
   && \left. -\Phi
   \left(-1,1,\frac{1}{q}\right)\right],
\eea
which is the first (massless) term of Eq.~\eqref{Pfermion}. For the massless bosons, the factor $(-1)^{s+1}$ does not arise, and the summation
yields Eq.~\eqref{Pboson}. Energy density can be obtained in a similar way, too.
\section{Pressure for the massive case: up to $\mathcal{O}(m^2T^2)$}
The pressure of a gas of massive fermions is given by,
\bea
P= \frac{g}{6\pi^2} \int_0^{\infty} dp 
\frac{p^4 (p^2+m^2)^{-\frac{1}{2}}}{\left[\left(1+(q-1)\frac{\sqrt{p^2+m^2}}{T}\right)^{\frac{q}{q-1}}+1\right]}
\label{Pmassive}
\eea
To find out the pressure of a gas of massive fermions up to $\mathcal{O}(m^2T^2)$, we first
expand the integrand in Eq.~\eqref{Pmassive} in a series of mass $m$, and retain only the
$\mathcal{O}(m^2T^2)$ term. Hence, the pressure can be approximated as,

\bea
P \approx P^{(0)} + P^{(2)},
\eea
where $P^{(0)}$ is the massless term and $P^{(2)}$ is the $\mathcal{O}(m^2T^2)$ term.
$P^{(0)}$ can be evaluated using the steps detailed in the previous section. 
When $P$ is expanded in a series, $P^{(2)}$ is given by,
\bea
P^{(2)} &=& \frac{g m^2}{12\pi^2} \int_0^{\infty} dp 
\left[
   -
   \frac{ p}{\left(\frac{p
   (q-1)}{T}+1\right)^{\frac{q}{q-1}}+1} \right. \nn\\
&&\left.   -
   \frac{ p^2 q \left(\frac{p
   (q-1)}{T}+1\right)^{\frac{q}{q-1}-1}}{T
   \left[\left(\frac{p
   (q-1)}{T}+1\right)^{\frac{q}{q-1}}+1\right]^2}  \right].
\label{POm2}
\eea
The first term on the r.h.s of Eq.~\eqref{POm2} contains a single power of the Tsallis Fermi-Dirac distribution and can be evaluated 
following the steps detailed in the previous section. 

The second term, which contains a square of the Tsallis distribution, can be expressed 
in terms of a partial derivative of the Tsallis distribution with respect to momentum $p$. 
Then carrying out the integration by parts, one may be able to express it in terms of 
a single power of the Tsallis distribution. Afterwards, following the steps already discussed,
one may arrive at the expression given by the second term in Eq.~\eqref{Pfermion}.

\end{appendix}

%
%
%
%
%
%


%
%

\end{document}